\documentclass[12pt]{article}

\usepackage{amsmath}
\usepackage{amssymb}
\usepackage{mathtext}
\usepackage{graphicx}

\begin{document}
\title{Motion of vortex sources on a plane and a sphere}
\author{Alexey V. Borisov, Ivan S. Mamaev\\
Institute of Computer Science, Udmurt State University\\
1, Universitetskaya str., 426034 Izhevsk, Russia\\
Phone/Fax: +7-3412-500295\\
E-mail: borisov@rcd.ru, mamaev@rcd.ru}
\maketitle
\begin{abstract}
The Equations of motion of vortex sources (examined earlier by Fridman
and Polubarinova) are studied, and the problems of their being Hamiltonian
and integrable are discussed. A system of two vortex sources and three
sources-sinks was examined. Their behavior was found to be regular.
Qualitative analysis of this system was made, and the class of Liouville
integrable systems is considered. Particular solutions analogous to the
homothetic configurations in celestial mechanics are given.
\end{abstract}


Keywords: Point vortex, sink/sourse, particular solution, Poisson
structure.

\newpage
A point vortex model going back to G.\,Kirchhoff is used to describe many
effects in classical hydrodynamics. Various problems solved with the use
of this model are reviewed in \cite{bib5}. Less known is a more general
problem formulation which describes the motion of more complex
singularities in a fluid (these singularities combine vortical properties
and properties of a system of sources and sinks). This model is sometimes
of greater use for hydrometeorology. It was first obtained by
A.\,A.\,Fridman, the famous Russian scientist who worked in hydrodynamics
and cosmology, and his disciple  P.\,Ya.\,Polubarinova (known also as
Polubarinova-Kochina, April 13, 1899 -- July 13, 1999; she was
academician N.\,E.\,Kochin's wife). In \cite{bib1}, the motion of two
sources-sinks is also studied and general integrals of motion equations
are examined. Hamiltonian equations of motion of sources--sinks were
studied by A.\,A.\,Bogomolov \cite{bib2} (this work \cite{bib2} performed
independently of \cite{bib1}). The main features in the motion of
sources and sinks (hereafter, they will be referred to as sources) are
used for modeling heat convection in a flat horizontal fluid layer, for
example, periodic convection cells. The system of two sinks was integrated
in \cite{bib1}; it was mentioned that the case $N\ge3$, as well as the
case of the equations of motion of $N$ point vortices and $N$ point
masses, causes fundamental difficulties. Among the studies dealing with the
dynamics of sources, we should also mention \cite{bib8,bib9}, where
chaotic advection of a fluid in a field of pulsating sources is examined,
and a number of examples in molecular biology are given.

\section{Equations of motion of vortex sources}

It is well known~\cite{Kirh} that the dynamics of point vortices in a
plane (vortex lines) in an ideal incompressible fluid can be described by
the (Hamiltonian) system of first-order equations
\begin{equation}
\label{eq_1}
\dot{\bar{z}}_k^{} = \frac{1}{2\pi} \sum_{l \ne k}^n \frac{i
\Gamma_k^{}} {z_k^{} - z_l^{}},
\end{equation}
where $z_k^{} = x_k^{} + i y_k^{}$ are
coordinates of vortices in the complex form, and $\Gamma_k^{}$ are their
intensities. The velocity of fluid~$v = v_x + iv_y^{}$ at a given point of
the plane is determined by the expression
\begin{equation}
\label{eq_2}
 \bar{v}(z,t) =
\frac{1}{2\pi} \sum_{k=1}^n \frac{i \Gamma_k^{}}{z - z_k^{}(t)},
\end{equation}
so that the velocity near a fixed vortex $z_k^{}$ can be represented in the
form of the series
\begin{equation}
\label{eq_3}
 \bar{v}(z) = \frac{i \Gamma_k/2\pi}{z -
z_k^{}} + \sum_{s = 0}^\infty a_s^{} (z - z_k)^s.
\end{equation}

Thus, the dynamics of point vortices is the dynamics of singularities in
the field of an ideal incompressible fluid of a special kind~\eqref{eq_3}.
In their work mentioned above, A.\,A.\,Fridman and P.\,Ya.\,Polubarinova
\cite{bib1} consider the dynamics of singularities of a more general form
(these singularities persist during fluid motion), in the case when the
expansion of velocity near the singularity is determined by the Laurent
series
\begin{equation}
\label{eq_4}
\bar{v}(z) = \sum_{s=1}^\infty \frac{b_s^{}}{(z -
z_k^{})^s} + \sum_{s=0}^{\infty}a_s(z - z_0^{})^s,
\end{equation}
where $a_s^{}$, $b_s^{}$ are complex numbers such that all $b_s^{}$ do not vanish
simultaneously.

To determine the velocity of the singularity we use the following
principle~\cite{bib1}:
\begin{itemize}
\item[] Suppose that~$K$ is a ring near a singularity with the inner
radius~$l_1$ and external radius~$l_2$ (see~figure~1); if
the fluid bounded by the ring~$K$, solidify in the moment considered
(without redistribution of the surrounding velocity field), the obtained
buoy will have a translational velocity $v_0$ and the angular velocity
$\omega_0$. When the radii $l_1,\,l_2$ tend to zero, we obtain the
respective {\it velocities of the singularity}.
\end{itemize}

Using~\eqref{eq_4} and taking into account the conservation laws of
momentum and angular momentum, we obtain the following:
\begin{equation}
\label{eq5}
v_0(l_1,l_2)=a_0,\quad \omega_0(l_1,l_2)=-\frac{2\text{Im } b_1}{l_1^2+l_2^2},
\end{equation}
and passage to the limit yields
\begin{equation}
\label{eq6}
v=a_0,\quad \omega=
\left\{
\begin{aligned}
&\infty, && \text{if  } \text{Im } b_1 \neq 0,\\
&0, && \text{if  } \text{Im } b_1=0.
\end{aligned}
\right.
\end{equation}

Thus, the translational velocity of the singularity (in the complex form)
is equal to the absolute term in the expansion of flow velocity into
Laurent series, and the angular velocity does not vanish only if $\text{Im } b_1
\neq 0$, which corresponds to an ordinary point vortex in a fluid with
$\omega=\infty$.

{\bf Remark 1.}
{\it The above reasoning shows that additional physical considerations (such as
the existence of minimum admissible sizes within which the flow has
no singularities and the like), should be involved in the construction of
dynamic models of higher-order singularities (dipoles, quadrupoles, and
the like) and the evaluation of their angular velocity.}

The simplest singularities (which also are isotropic, that is, their
angular velocity has no effect on their translational velocity) are the
following first-order singularities (i.e., $b_i=0, \, i>2$):
\begin{itemize}\itemsep-0.3pt
\item[1.] vortex $(\text{Im } b_1 \neq 0,\, \text{Re } b_1=0)$;
\item[2.] sources $(\text{Im } b_1=0,\, \text{Re } b_1 \neq 0)$;
\item[3.] vortex source $(\text{Im } b_1 \neq 0,\, \text{Re } b_1 \neq 0)$.
\end{itemize}

The equation of vortex motion can be written as
\begin{equation}
\frac{d\bar{z}_k}{dt} = \sum_{k \ne l}^n \frac{\mathrm{A}_l}{z_k - z_l},\quad \mathrm{A}_l^{} =
\frac{\mathrm{K}_l + i \Gamma_l}{2\pi},\quad k=1,\ldots,n,\label{kappa}
\end{equation}
where $\mathrm{K}_l$ and $\Gamma_l$ are the intensities of the source and
the vortex, respectively. A particular case of the system~\eqref{kappa} is
the system of equations for motion of vortices~\cite{Kirh} and
sources~\cite{bib1,bib2}, which, as was found, can be represented in
Hamiltonian form.

\section{Invariance of motion equations; integrals and Hamiltonian character}
At arbitrary complex intensities $\mathrm{A}_k$, equations~\eqref{kappa}
are invariant with respect to the group of motions of the plane $E(2)$. The
invariance with respect to translations results in the existence of two
linear integrals, which in a complex form can be written as
\begin{equation}
\label{eq_8}
 Z
= \sum_{k=1}^n \mathrm{A}_k \bar{z}_k = \text{const}.
\end{equation}

In the case of ordinary vortices and sources, we have the well-known
integrals for the vorticity center and the center of divergence:
\begin{equation}
\label{eq6.5}
\begin{aligned}
\text{(for vortices)} \quad Q_v^{} = \textstyle \sum \Gamma_k
x_k = \text{const},\quad P_v^{} = \sum \Gamma_k y_k^{}
= \text{const},\\
\text{(for sources)} \quad Q_s^{} = \textstyle \sum \mathrm{K}_k x_k =
\text{const},\quad P_s^{} = \sum \mathrm{K}_k y_k^{} = \text{const}.\\
\end{aligned}
\end{equation}
In the general
case, $Z$ in \eqref{eq_8} determines {\it the center of intensities}.
Clearly, if $\sum A_k \neq 0$, we can choose the origin of the coordinates
such that $Z{=}0$.

Unlike the vortex dynamics, the rotational symmetry, which corresponds to
a field of symmetries of the type of $\hat{\boldsymbol {v}}=\sum\limits_{k=1}^{n}(x_k{\frac{\partial}{\partial{x_k}}}-
y_k{\frac{\partial}{\partial{x_k}}})$, does not result, in the general
case, in the existence of an additional integral. In this case, we have
only a differential relation
\begin{equation}
\label{eq_10}
\sum_k\mathrm{A}_k^{} z_k \dot{\bar{z_k^{}}} = \sum_{k,l} \mathrm{A}_k
\mathrm{A}_l.
\end{equation}
In the case of vortices~\eqref{eq_10}, this relationship allows us to
obtain the well-known integral of momentum
$$ I_v = \sum_{k=1}^n
\Gamma_k (x_k^2 + y_k^2)=\text{const},
$$
and in the case of sources, we obtain
$$
\sum \mathrm{K}_k^{} (\dot x_{k} y_k - x_k
\dot{y_k^{}}) = 0,\quad \sum \mathrm{K}_k^{} (\dot x_{k} x_k + y_k
\dot{y_k^{}}) = \sum_{k,l} \mathrm{K}_l^{} \mathrm{K}_k^{}.
$$
Thus, in the case of sources, we have the non-autonomous integral
\begin{equation}
\label{eq11}
 I_s = \sum {\mathrm{K}_k}(x_k^2 + y_k^2) - 2\big(\sum_{k,l} \mathrm{K}_k
\mathrm{K}_l\big)t = \text{const}.
\end{equation}
It is clear that, when $\sum_{k,\,l}
\mathrm{K}_k \mathrm{K}_l = 0$, in the case of sources, we obtain an
ordinary autonomous integral. In the general case of vortex sources,
~\eqref{eq_10} fails to yield even a non-autonomous integral.

It can be easily shown that the divergence of the right-hand part of
equations~\eqref{kappa} is equal to zero; therefore, we have

{\bf Proposition 1}
{\it The Equations of motion of vortex sources conserve the standard (i.e.,
constant-density) invariant measure.}

In addition to the above-mentioned integrals and relations, equations
\eqref{kappa} in particular cases of vortices and sources yield one more
integral $H$ and also can be written in the Hamiltonian form:
$$
\dot x_i^{} = \{x_i,H\}, \quad \dot y_i = \{y_i, H\}.
$$

This representation is well known for vortices; the Hamiltonian and the
Poisson bracket are:
\begin{equation}
\label{eq12}
H = -\frac{1}{4\pi} \sum_{i < j}
\Gamma_i \Gamma_j \ln|z_i - z_j|^2,\quad \{x_i, y_j\}=\Gamma_i^{-1}\delta_{ij}.
\end{equation}

Analogous representation for sources was suggested in~\cite{bib2}:
\begin{equation}
\label{eq13}
 H = \frac{1}{2\pi} \sum_{k < l} \mathrm{K}_k \mathrm{K}_l
\arctan \frac{y_k - y_l}{x_k - x_l}, \quad  \{x_i^{}, y_j^{}\} =
\mathrm{K}_k^{-1} \delta_{ij}.
\end{equation}
The integral \eqref{eq13} was obtained in \cite{bib1}. As is the case with
vortices, the integrals of motion \eqref{eq6.5} and~\eqref{eq11} are
non-involutive, and satisfy the following commutation relations:
\begin{equation}
\label{eq11.5}
 \{Q_s^{},P_s^{}\} = \sum_{i} {\rm K}_i^{}, \quad
\{I_s^{},Q_s^{}\} = -2P_s,\quad \{I_s^{}, P_s^{}\} = 2Q_s.
\end{equation}

Equations~\eqref{eq11} and \eqref{eq11.5}, by virtue of the Liouville
theorem, imply the following conclusion.

{\bf Proposition 2}
{\it If  in the problem of motion of three sources $\sum_i \mathrm{K}_i = 0$
or~$\sum_{i,j} \mathrm{K}_i \mathrm{K}_j = 0$, the system is completely
integrable in these cases.}

Indeed, if $\sum_i\mathrm{K}_i = 0$, we have three involutive integrals,
for example, $Q_s^{}$, $P_s^{}$, and $H$, while if $\sum_{i,j}
\mathrm{K}_i \mathrm{K}_j = 0$, such integrals are $I_s^{}$, $Q_s^2 +
P_s^2$, $H$.

Moreover, for the case of four sources, we obtain

{\bf Proposition 3}
{\it Suppose that in the problem of four sources, we have
$\sum_{i}\mathrm{K}_i = \sum_{i,j} \mathrm{K}_i \mathrm{K}_j = 0$, then
the system \eqref{kappa} is completely integrable on the invariant
manifold determined by condition $Q_s^{} = P_s^{} = 0$.}

Indeed, in this case we have three autonomous independent integrals
$Q_s^{}$, $P_s^{}$, and $I_s^{}$, which are in involution on this
manifold. Note that, for the problem of four vortices to be integrable, it
is sufficient to require that $Q_s=P_s=0$, $\sum\Gamma_i=0$.

{\bf Remark 2}
{\it In the general case of vortex sources, we can consider the complex
function}
$$
\mathcal{H} = \sum \mathrm{A}_k^{} \mathrm{A}_l^{} \ln (z_k - z_l),
$$
which satisfies the following relations~\cite{bib1}:
$$
\mathrm{A}_k^{} \frac{d \bar{z_k^{}}}{dt} = \frac{\partial{\mathcal{H}}}{\partial z_k^{}},\quad
\bar{\mathrm{A}}_k^{} \frac{d {z_k^{}}}{dt} = \frac{\partial{\bar{\mathcal{H}}}}{\partial \bar{z}_k^{}}.
$$

\section{Motion of two vortex sources}
This problem is completely integrable by quadratures, and its particular cases
are well known.
\subsection*{a) Two vortices~\cite{81}}
If $\Gamma_1 + \Gamma_2 \ne 0$, both vortices move along concentric
circles around a center of vorticity with coordinates
$\frac{\Gamma_1 z_1 + \Gamma_2 z_2}{\Gamma_1 + \Gamma_2 }$,
remaining on a common line with this center, and the angular rotation
velocity is

$\omega = \frac{\Gamma_1 + \Gamma_2}{2 \pi d^2}$, where $d$ is the
distance between the vortices.

If $\Gamma_1 + \Gamma_2 = 0$, the motion of vortices is translational in
the direction perpendicular to the straight line connecting them. And the
velocity of vortices is $v = \frac{\Gamma_1}{2\pi d}$.

\subsection*{b) Two sources~\cite{bib1,bib2}}

If $\mathrm{K}_1 + \mathrm{K}_2 \ne 0$, the sources move along a single
straight line fixed in space and approach the center of intensity with
coordinates $\frac{\mathrm{K}_1 z_1 + \mathrm{K}_2 z_2}{\mathrm{K}_1 +
\mathrm{K}_2}$; and the square of the distance between them varies in
accordance with the following formula $d^2 = d_0^2 + 2 (\mathrm{K}_1 +
\mathrm{K}_2)t$.

If $\mathrm{K}_1 + \mathrm{K}_2 = 0$, the sources move along a fixed line
so that the distance between them remains constant and their velocity is
$v = \frac{\mathrm{K}_1}{2\pi d}$.

\subsection*{c) Two arbitrary vortex sources}

The equations of motion in this case are
\begin{equation}
\label{eq_15}
\dot{\bar{z}}_1 = \frac{\mathrm{A}_2}{z_1 - z_2},\quad \dot{\bar{z}}_2 =
\frac{\mathrm{A}_1}{z_1 - z_2}.
\end{equation}
Suppose that $\mathrm{A}_1 +
\mathrm{A}_2 \ne 0$; without loss of generality, we can assume that the
center of intensity \eqref{eq_8} lies in the origin of coordinates, that
is
$$
\bar{\mathrm{A}}_1 z_1 + \bar{\mathrm{A}}_2 z_2 = 0.
$$
Expressing from here the coordinates of each particle and passing to polar
coordinates $z_i^{} = \rho_i e^{i \varphi_i^{}}$, $i = 1,2$, we obtain
$$
\rho_k \dot \rho_k^{} = a_k^{},\quad \rho_k^2 \dot \varphi_k = -
b_k^{},\quad k = 1,2,
$$
where $a_k^{} + i b_k^{} = (\bar{A}_1 +
\bar{A}_2)^{-1}|A_l|^2$, $k \ne l$. Solving these equations yields the law
of motion:
\begin{equation}
r_i^2 = C_i^{(0)} + 2a_i^{} t, \quad \varphi_i^{} =
\varphi_{i}^{(0)} - \frac{b_i}{2a_i}\ln (C_i^{(0)} + 2 a_i^{}t),
\end{equation}
where
$\varphi_i^{(0)},\,C_i^{(0)}$ are constants of integration. Thus,
equations \eqref{eq_15} are integrable by quadratures.

{\it In this case, the trajectories of vortex sources are spirals with an
infinite number of turns around the center of intensities}.

If $\mathrm{A}_1 + \mathrm{A}_2 = 0$, assuming that $Z =
\bar{\mathrm{A}}_1 z_1 + \bar{\mathrm{A}}_2 z_2$, we find that
$$
z_2 = z_1 - Z/\bar{\mathrm{A}}_1, \quad \dot{\bar{z}}_1 = -
\frac{{|\mathrm{A}_1|}^2}{Z} =\text{const},
$$
that is, the vortex sources move
with a constant velocity along parallel straight lines at an angle of
$\alpha$\, $(\tan{\alpha}=\frac{\Gamma_1}{K_1})$ with respect to the segment
connecting them.

Interestingly, equations \eqref{eq_15} also have an integral of the energy
type, which is a superposition of integrals \eqref{eq12} and
\eqref{eq13}:
\begin{equation}
\label{eq17}
H = \frac{(\Gamma_1 + \Gamma_2)}{4\pi}\ln|z_1 -
z_2|^2 + \frac{(\mathrm{K}_1 + \mathrm{K}_2)}{2\pi} \arctan \frac{y_1 -
y_2}{x_1 - x_2}.
\end{equation}

It is well known that the presence of three integrals in a
four-dimensional system (in this case, these are integrals $X = \text{Re}
(\bar{\mathrm{A}}_1 z_1 + \bar{\mathrm{A}}_2 z_2), Y = \text{Im}
(\bar{\mathrm{A}}_1 z_1 + \bar{\mathrm{A}}_2 z_2)$ and~$H$) implies in
that such system can be represented in the Hamiltonian form~\cite{bib3}.
If we take the function~$H$ in~\eqref{eq17} as the Hamiltonian, the respective
Poisson brackets are
\begin{equation}
\begin{aligned}
\{f,g\}dx_1 \wedge
dy_1 \wedge dx_2 \wedge dy_2 = \ae\mbox{ } df \wedge dg \wedge dX \wedge dY,\\
 \ae = \left({(\mathrm{K}_1+\mathrm{K}_2)}^2
+{(\Gamma_1+\Gamma_2)}^2\right)^{-1}.
\end{aligned}
\end{equation}
Explicitly, we obtain
\begin{equation}
\begin{gathered}
\{x_i,y_i\} = \ae(\mathrm{K}_j^2 + \Gamma_j^2),\quad \{x_i, y_j\} = - \ae
(\mathrm{K}_i \mathrm{K}_j +
\Gamma_i \Gamma_j),\\
\{x_i, x_j\} = \{y_i, y_j\} = \ae (\mathrm{K}_i \Gamma_j - \mathrm{K}_j
\Gamma_i),\quad i \neq j.
\end{gathered}
\end{equation}

{\bf Remark 3} {\it Generally speaking, the system \eqref{eq_15} can be represented in the
Hamiltonian form in three different ways, but the Poisson brackets thus
obtained are not constant.}

{\bf Remark 4} {\it
Although the system is Hamiltonian, the merging of vortex source (at
$a_i^{} < 0$) within finite time is not excluded.}

\section{Three-sources system}
\subsection{Reduction}
It is common knowledge that a three-vortex system on a plane is Liouville
integrable and its phase space in the compact case is foliated into
two-dimensional tori filled with quasiperiodic motions. To study that problem methods of
qualitative analysis of Hamiltonian systems (geometric interpretation,
construction of bifurcation diagrams, and so on; see \cite{bib5,bib7},
where a vast bibliography is presented) can be used.

In the general case, the three-source problem is not Liouville integrable
by quadratures. Its behavior, however, is regular. Indeed, this problem is
determined by a Hamiltonian system with the Poisson bracket $\{x_i, y_j\} =
\mathrm{K}_i^{-1} \delta_{ij}$ and the Hamiltonian
\begin{equation}
\label{t1}
 H = \frac{1}{2\pi} \sum_{i < j}^3 \mathrm{K}_i \mathrm{K}_j
 \theta_{ij},\quad
\theta_{ij} = \arctan \frac{y_i - y_j}{x_i - x_j}.
\end{equation}

As in the general case, the equations of motion of this system are
invariant with respect to the $E(2)$ group (group of motions of the
plane). According to the general theory~\cite{bib3},
we obtain a closed system for the invariants of the symmetry group. In
our case (as in the vortex theory), it is convenient to take distances
between the sources $M_i = |z_j - z_k|^2$, $i \ne j \ne k \ne i$ as
invariants. We obtain
\begin{equation}
\label{t2}
\dot{M}_k^{} = 2 \big(\sum_l
\mathrm{K}_l^{}\big) + \mathrm{K}_k \big(M_k^{} (M_i^{-1} + M_j^{-1}) -
M_i M_j^{-1} - M_j M_i^{-1}\big).
\end{equation}

Note that such reduction into invariants of the group of symmetries in
this case differs from the traditional Hamiltonian reduction. This is due
to the fact that system~\eqref{t1} does not admit autonomous integrals, the
Hamiltonian vector field of rotations is associated with the nonautonomous
integral \eqref{eq11}, which linearly varies with time.

In the Hamiltonian form, reduction only by one degree of freedom is
possible (based on integrals~\eqref{eq_8}). In this case, the
distances $M_i$ should be complemented by one of the angles $\Theta
=\theta_{12}$; the Hamiltonian of the reduced system and the Poisson bracket
can be written as:
$$
\begin{gathered}
H = \left(\sum_{i < j} \mathrm{K}_i \mathrm{K}_j\right)\Theta -
\mathrm{K}_2 \mathrm{K}_3 \arccos \left(\frac{M_3 + M_1 - M_2} {2
\sqrt{M_1 M_3}}\right) + \mathrm{K}_1 \mathrm{K}_2 \arccos \left(\frac{M_3
+ M_2 - M_1}{2 \sqrt{M_2
M_3}}\right),\\
\{M_i^{}, M_j^{}\} = \varepsilon_{ijk} \mathrm{K}_k^{-1} \Delta, \quad \{M_3, \Theta\}
= 2 (\mathrm{K}_1^{-1} +
\mathrm{K}_2^{-1}),\\
\{M_2^{}, \Theta\} = \mathrm{K}_2^{-1} M_3^{-1} (M_3 + M_1 - M_2),\quad \{M_2,
\Theta\}= \mathrm{K}_1^{-1} M_3^{-1} (M_3 + M_2 - M_1),
\end{gathered}
$$
where $\Delta = \text{Im} (z_1 - z_3)(\overline{z}_2 - \overline{z}_3)$ is the doubled
oriented area of the triangle formed by the sources; this area can be
expressed in terms of the distances by the Heron formula:
\begin{equation}
 4 \Delta^2 = 2\big(M_1 M_2 + M_2 M_3 + M_3 M_1 \big) - M_1^2 - M_2^2 -
M_3^2.\label{t18}
\end{equation}
The equation
\begin{equation}
 \dot \Theta = \mathrm{K}_3 \frac{\Delta
(M_1 - M_2)}{M_1 M_2 M_3},\label{d19}
\end{equation}
 which describes the evolution
of~$\Theta$, completely determines the absolute dynamics of the sources
(the remaining angles $\theta_{23}, \theta_{13}$ can be found from the law of
cosines). Thus, the Hamiltonian reduction in this case is a simple
restriction on the level of integrals \eqref{eq_8}.

Let us consider equations \eqref{t2} in more detail. Direct calculations
readily show that system \eqref{t2} also has the nonautonomous integral
\begin{equation}
\label{t3}
\sum_{i < j} \mathrm{K}_i \mathrm{K}_j M_k - 2 \left(\sum
\mathrm{K}_k\right)\left(\sum_{i < j} \mathrm{K}_i \mathrm{K}_j\right)t =
\text{const}
\end{equation} and conserves the invariant measure with the density
$$
\rho = (2 \Delta)^{-1} = \left(2 M_1 M_2 + 2 M_2 M_3 + 2 M_3 M_1 - M_1^2 - M_1^2 -
M_3^2\right)^{-1/2}.
$$
Equations~\eqref{t2} are homogenous so they also admit the one-parameter
symmetry group \mbox{$M_i \to M_i$}, $t \to \lambda t$, which allows an
additional reduction of the order by a change in the variables and the
time in the following form (the so-called projection
procedure):
$$
M_1 = x M_3,\quad M_2 = y M_3,\quad dt = M_3 x y d \tau.
$$
We have
\begin{equation}
\begin{aligned}
x' = \frac{dx}{d\tau} = x \bigl((\mathrm{K}_1 -
\mathrm{K}_3)y^2 - \mathrm{K}_3 x^2 + (\mathrm{K}_1 + 2 \mathrm{K}_2 + 2
\mathrm{K}_3)
xy - (\mathrm{K}_1 - \mathrm{K}_3)x - \\
 - (2\mathrm{K}_1 + 2\mathrm{K}_2 + \mathrm{K}_3)y + \mathrm{K}_1\bigr),\\
y' = \frac{dy}{d \tau} = y \bigl((\mathrm{K}_2 - \mathrm{K}_3)x^2 - \mathrm{K}_3 y^2 +
(2\mathrm{K}_1 + \mathrm{K}_2 + 2
K_3)xy - (2\mathrm{K}_1 + 2\mathrm{K}_2 + \mathrm{K}_3)x- \\
 - (\mathrm{K}_2 - \mathrm{K}_3)y + \mathrm{K}_2\bigr).\label{t4}
\end{aligned}
\end{equation}

Thus, qualitative analysis of the system considered here reduces to the
examination of the system on the plane~\eqref{t4}. As a result, the
dynamics of the system \eqref{t1} is regular, although it does not have a
complete set of global integrals.

\subsection{Homothetic configurations}

In vortex dynamics, an important role in the qualitative analysis of
systems played by stationary configurations, with which stationary
points of the reduced system were associated. {\it Homothetic
configurations} (this terminology was adopted from celestial mechanics),
i.e., configurations that remain self-similar in any time moment, play an
analogous role in the dynamics of sources. This condition can be
represented in the following form
\begin{equation}
\label{t5}
{(M_i M_j^{-1})}\,\dot{} = 0, \quad i,j = 1,2,3.
\end{equation}
{\bf Remark 5}
{\it It can be readily seen that, because of the existence of integral
\eqref{t3}, the system of equations \eqref{t2} has no fixed points.}

The following simple statement is true:

{\bf Proposition 4}
{\it The only possible homothetic configurations in a three-source system are}:
\begin{enumerate}
\item {\it equilateral configuration, in which the sources form an equilateral triangle}:
$M_1 = M_2 = M_3$; {\it the vortices move along fixed straight lines, which
cross in the center of intensity, and the equality}
$$
\dot M_i = 2\sum_l \mathrm{K}_l
$$
{\it is valid;}

\item {\it collinear configurations, such that the vortices lie on the single fixed
straight line} $M_1 = z^2 M_3$, $M_2 = (1 - z)^2 M_3$, {\it where} $z$
{\it is a root of the  cubic equation}
\begin{equation}
\label{t6}
p(z) = (\mathrm{K}_1 + \mathrm{K}_2)z^3 - (2\mathrm{K}_1 + \mathrm{K}_2)z^2 -
(\mathrm{K}_2 + 2\mathrm{K}_3)z + \mathrm{K}_2 + \mathrm{K}_3 = 0,\vspace{-1mm}
\end{equation}
{\it such that}
$$
\dot M_3 = 2 \frac{(\mathrm{K}_1 + \mathrm{K}_2)z(z-1) - \mathrm{K}_3}{z(z-1)} = \text{const}.
$$
\end{enumerate}

PROOF.

According to~\eqref{t5}, homothetic configurations are fixed points of the
system~\eqref{t4}. The cases where $x=0$ or $y=0$ correspond to
singularities of the initial system, where one of the distances vanishes;
therefore, they should be left out of consideration. The remaining fixed
points can be found in the following way. Note that equation $\mathrm{K}_3
\frac{x'}{x} + (\mathrm{K}_1 - \mathrm{K}_3)\frac{y'}{y}=0$ is linear with
respect to $y$; solving this equation with respect to $y$ and substituting
$x' = 0$, we obtain, after cancellation, the equation that determines a
homothetic configuration
\begin{multline}
(x-1)\big((\mathrm{K}_1 + \mathrm{K}_2)^2 x^3 - (4\mathrm{K}_1^2 + 3
\mathrm{K}_2^2 + 6 \mathrm{K}_1 \mathrm{K}_2 +
4 \mathrm{K}_1 \mathrm{K}_3 + 4 \mathrm{K}_2 \mathrm{K}_3)x^2+\\
+(3 \mathrm{K}_2^2 + 4 \mathrm{K}_3^2 + 4 \mathrm{K}_1 \mathrm{K}_3 + 4
\mathrm{K}_1 \mathrm{K}_2 + 6 \mathrm{K}_2 \mathrm{K}_3)x - (\mathrm{K}_2
+ \mathrm{K}_3)^2\big)=0.\label{eq_22}
\end{multline}

We find from here one more solution $x=1$, which, as can be readily seen,
corresponds to $y=1$; that is, this solution corresponds to an equilateral
configuration. The remaining equation \eqref{eq_22}, which is cubic with
respect to $x$, with the substitution $x = z^2$ can be represented in the
form
$$
p(z)p(-z) = 0,
$$
where $p(z)$ is the polynomial \eqref{t6}.
Substituting the obtained solution for $x$ in the equation for $y$ yields
$y = (1 \pm z)^2$, that is, this solution corresponds to a collinear
configuration (with differently ordered sources).

Using equation~\eqref{d19}, we find that in both cases, $\dot{\Theta}=0$,
(by virtue of either $M_1{=}M_2$ or $\Delta=0$), that is, the sources move
along straight lines fixed in the space. The evolution rates of the
squared distances can be immediately found from equations \eqref{t2} after
the substitution of the obtained solutions.\mbox{   }$\blacksquare$

\subsection{Geometrical interpretation and qualitative analysis}

It is well known that complete qualitative analysis can be made for both
the motion of a reduced system and the absolute motion \cite{bib7}. A
convenient tool for study in this case is a projection of the phase
flow on the plane of the integral of momentum in the space of mutual
distances $M_i,~i=1,2,3$. We will show that, in the case of three sources,
such projection also exists; the corresponding plane system admits
complete qualitative analysis and gives a complete description of both
relative and absolute dynamics of sources.

Let us write a nonautonomous integral of the system \eqref{t2} in the form
\begin{equation}
\label{g1}
D = 2(\sum \mathrm{K}_i)(\sum \mathrm{K}_i^{-1})t = \text{const},\quad
D = \sum \mathrm{K}_i^{-1} M_i
\end{equation}
and define the projection and time
transformation by the formulas
$$
 x_i^{} = \frac{M_i}{D},\quad d\tau =
\frac{dt}{D}, \quad i=1,2,3.
$$
 The equations of motion for the projective
variables are
\begin{equation}
\label{g2}
\begin{gathered}
x_k' \!=\! 2 \sum \mathrm{K}_i \!+\! \mathrm{K}_k (x_k(x_i^{-1} \!+\!
x_j^{-1}) \!+\! x_i^{-1}x_j \!-\! x_j^{-1}x_i) \!-\! 2\left(\sum
\mathrm{K}_i\right)\left(\sum \mathrm{K}_i^{-1}\right)x_k,\\
k \ne i \ne j \ne k;
\end{gathered}
\end{equation}
clearly, the trajectories of the system~\eqref{g2} that correspond to
the trajectories of the original system lie on the invariant manifold
\begin{equation}
\label{g3}
\sum_{i=1}^3 \mathrm{K}_i^{-1} x_i = 1.
\end{equation}
This manifold is the
plane in the three-dimensional space $x_1$, $x_2$, $x_3$.

However, imply that the Heron formula \eqref{t18} and the conditions
$M_i > 0$, the domain of possible motions lies within one half of the cone
determined by the inequality
\begin{equation}
\label{g4} 2(x_1x_2 + x_2x_3 + x_3x_1) - x_1^2
- x_2^2 - x_3^2 \ge 0.
\end{equation}

Thus, the boundary of the domain of possible motions on the plane
\eqref{g3}, which is determined by inequality \eqref{g4}, is a
second-order curve of one of the three types:

\begin{enumerate}
\item[1)] ellipse (if $(\sum \mathrm{K}_i)/(\mathrm{K}_1 \mathrm{K}_2 \mathrm{K}_3) > 0$);
\item[2)] hyperbola (if $(\sum \mathrm{K}_i)/(\mathrm{K}_1 \mathrm{K}_2 \mathrm{K}_3) < 0$);
\item[3)] parabola (if $\sum \mathrm{K}_i = 0$).
\end{enumerate}

A one-to-one mapping exists between the trajectories of the system
\eqref{g2} in the domain considered here and the trajectories of the
reduced system of three sources \eqref{t2}. Thus, we have constructed a
projection of the phase flow of a three-source system, which is completely
identical to the analogous projection in the vortex dynamics (up to the
change $\mathrm{K}_i \to \Gamma_i$) \cite{bib5, bib7}.

Let us consider in more detail the structure of the phase portrait in the
first case (see figures~2,\,3), which, by analogy with vortex dynamics
will be referred to as {\em ``compact''}; clearly, in this case all
ratios $M_i/M_j$ remain bounded for all time. The second
({\em ``noncompact''}) case can be studied in a similar manner; however,
the latter, third case (as well as the case $\sum \mathrm{K}_i^{-1} = 0$)
requires special consideration because, according to~\eqref{g1}, an
autonomous integral exists in this case.

The structure of the phase flow of the system on the plane~\eqref{g3} is
completely determined by its singularities, fixed points, and
separatrices connecting fixed points. Statement 1 makes it possible to
show that the system on the plane considered here has either four fixed
points, which we denote $A$, $B_1$, $B_2$, and $B_3$ (figure~2), or two
fixed points~--- $A$ and $B_1$ (figure~3). Point $A$ corresponds to an
equilateral homothetic configuration and lies within the domain (ellipse);
points $B_1$, $B_2$, and $B_3$ lie on the boundary of the domain and
determine collinear homothetic configurations (since it is clear that for
these points, $\Delta = 0$). Additionally, the system \eqref{g2} has (in the
domain considered) three singularities $C_1$, $C_2$, and $C_3$, which lie
on the boundary \eqref{g4} in the tangent points of planes $x_i = 0$; they
correspond to the situation when the pair of sources merges.

As in the vortex dynamics, in the compact case, we distinguish two
cases, in which the portraits qualitatively differ:
\begin{enumerate}
\item[$1^\circ$.] all intensities have the same sign;
\item[$2^\circ$.] two intensities are positive (negative), while the
third intensity is negative (positive), and $\mathrm{K}_1 + \mathrm{K}_2 +
\mathrm{K}_3 < 0 (> 0)$.
\end{enumerate}

In the case of $1^\circ$, as can be seen from figure~2, collinear
configurations are saddle points, and the equilateral configuration is a
node point (at $\mathrm{K}_i < 0$, it is a sink, while at $\mathrm{K}_i >
0$, it is a source), such that all the three separatrices of collinear
points enter into it. All the trajectories start (at $\mathrm{K}_i < 0$)
from singularities $C_1$, $C_2$, and $C_3$, do not go beyond the domain
limited by the separatrices of collinear points $B_1$, $B_2$, $B_3$, and
merge in the fixed point $A$. (At $\mathrm{K}_i > 0$, the direction of
motion along these trajectories should be reversed.)

In the case $2^\circ$, the only fixed point $B_1$ (corresponding to the
collinear configuration) is a source ($\mathrm{K}_i < 0$); the equilateral
configuration now corresponds to a saddle point. Three separatrices of the
saddle point end in the singularities $C_1$ $C_2$, $C_3$, respectively, and the fourth
separatrix, ends in the fixed point $B_1$. A general phase portrait
is given in figure~3.

\section{Homothetic configurations for $n$ sources}
Of great importance in the dynamics of $n$ point vortices are the simplest
periodic solutions, i.~e., stationary configurations such that vortices
rotate around a common center of vorticity without changing their mutual
arrangement. Analogous solutions were found in the dynamics of $n$ point
sources, with the only difference that the points move along fixed
straight lines passing through the center of divergence of the sources so
that their configuration remains self-similar at any time. Such
configurations will be referred to as {\em homothetic}.

Let us show that the following statement is true

{\bf Proposition 5} {\it
There exists a one-to-one map between the stationary
configurations of the dynamics and the homothetic configurations of the
dynamics of $n$ point sources.}

PROOF

Indeed, for any stationary configuration in a coordinate system
with the origin at the vorticity center, the coordinates of vortices are
as follows
$$
z_k(t) = \zeta_k e^{i \Omega t}, \quad k = 1,\ldots ,n,
$$
where $\Omega$ is the constant angular velocity of rotation of the
configuration, and the constant complex numbers $\zeta_k$ satisfy the
system of algebraic equations:
\begin{equation}
\label{ga1}
-2\pi \Omega \bar{\zeta_k} =
\sum_{l \ne k}^{n} \frac{\Gamma_l}{\zeta_k - \zeta_l},\quad k = 1\ldots n,
\end{equation}
 where $\Gamma_l$ are the intensities of vortices $\left(\mathrm{A}_l =
\frac{i \Gamma_l}{2 \pi}\right)$.

In the case of sources $(\mathrm{A}_l = \frac{\mathrm{K}_l}{2\pi})$, we
similarly find the particular solutions which specify the homothetic
configurations in the form
$$
z_k =\sqrt{c - 2\Omega t}\, \zeta_k,\quad k
= 1\ldots n,
$$
where $\zeta_k$ are complex numbers satisfying the system
of equations \eqref{ga1}, in which the change of variables $\Gamma_l \to
\mathrm{K}_l$ must be made.\mbox{    }$\blacksquare$

Note that the solution of the system of algebraic equations \eqref{ga1} is
not known, and the problem of finding the homothetic configurations of
sources also cannot be solved completely. A some of particular results
for stationary configurations in vortex dynamics are summarized in
\cite{bib7}.

{\bf Remark 6}
{\it In celestial mechanics, the homothetic configurations form a particular
case of a wider class of homographic configurations, when all points at
all times form such configurations, which rotate around a common
center of masses. There seems to be no such configurations, other than
homothetic, in the source dynamics. As far as we know, the question of
existence of such configurations in the case of vortex sources has not
been considered.}

\section{Vortex sources on the sphere}
Let us consider a possible generalization of the motion equations of
vortex sources (and other singularities in a fluid flow) for the case of a
sphere. To do this, we make a stereographic projection of the plane onto
the sphere and write the equations of vortex motion on the
sphere~\cite{bib5} in complex form.

In the case of the stereographic projection from the southern pole onto
the plane passing through the northern pole (see figure~4), the
rectangular coordinates of a point on the sphere can be written as
\begin{equation}
\begin{gathered}
x = \frac{\xi}{1+\lambda(\xi^2+\eta^2)},\quad
y = \frac{\eta}{1+\lambda(\xi^2+\eta^2)},\quad
z = R\frac{1-\lambda(\xi^2+\eta^2)}{1+\lambda(\xi^2+\eta^2)},\label{s1}
\end{gathered}
\end{equation}
where $\lambda={(2R)}^{-2}$ is the curvature.

Using the complex coordinates of vortices $\zeta_k = \xi_k + i\eta_k$, $k = 1\ldots n$ on
the plane considered, the motion equations of the vortices can be
represented in the form
\begin{equation}
\label{s2} \dot{\bar \zeta}_k =
\frac{\left(1+\lambda{|\zeta_k|}^2\right)^2}{2\pi} \left(\sum_{l\ne k}^n
\frac{i \Gamma_l}{\zeta_k - \zeta_l} - \left(\sum_{l\ne k}^n i
\Gamma_l\right) \frac{\lambda\bar{\zeta_k}}{\left(1 +
\lambda{|\zeta_k|}^2\right)} \right)\!.
\end{equation}

{\bf Remark 7}
{\it The factor before the second summand in \eqref{s2} in the book \cite{bib5}
appears to be erroneous.}

The velocity of the fluid determined by the vortices in any point of the
sphere is expressed analogous to~\eqref{s2}:
\begin{equation}
\label{s3} \dot{\bar \zeta} =
\frac{\left(1+\lambda{|\zeta_k|}^2\right)^2}{2\pi} \left(\sum_{k=1}^n
\frac{i \Gamma_l}{\zeta - \zeta_l} - \left(\sum_{l=1}^n i \Gamma_l\right)
\frac{\lambda\bar{\zeta}}{\left(1+\lambda{|\zeta_k|}^2\right)} \right)\!.
\end{equation}
It can be easily shown that the divergence of this velocity field is
zero and its vorticity is constant at any point of the sphere $\zeta \ne
\zeta_k$ and equals
\begin{equation}
\label{s35}
\omega = \frac{1}{4 \pi R^2} \sum_{k=1}^n\Gamma_k.
\end{equation}

By analogy with the planar case, let us replace the intensities $i
\Gamma_k$ by complex numbers $\mathrm{A}_k = \mathrm{K}_k + i \Gamma_k$;
we obtain the equations describing the motion of singularities. These singularities can
be interpreted as vortex sources on the sphere.
\begin{equation}
\label{s4}
\dot{\bar\zeta}_k = \frac{\left(1+\lambda{|\zeta_k|}^2\right)^2}{2\pi}
\left(\sum_{k=1}^n \frac{\mathrm{A}_k}{\zeta_k - \zeta_l} -
\left(\sum_{l\ne k} \mathrm{A}_l\right)
\frac{\lambda\bar{\zeta_k}}{\left(1+\lambda{|\zeta_k|}^2\right)}
\right)\!.
\end{equation}

Unlike the planar case, the velocity field determined by the singularities
\eqref{s4} (this field can be calculated by formula \eqref{s3} with
replacement $i\Gamma_l \to \mathrm{A}_l$), in addition to the vortex
sources in points $\zeta_k$, has a non-zero vorticity~\eqref{s35} and a
divergence equal to
\begin{equation}
\label{s5}
\kappa = - \frac{1}{4\pi R^2}\sum_{l = 1}^n
\mathrm{K}_l.
\end{equation}
at any point of the sphere $\zeta \ne \zeta_k$.

Clearly, such corrections in the motion equations of a fluid on a sphere
stem from the compactness of the sphere, and if a point source or a vortex
appears in some point, at the same time, a point sink or sinks uniformly
distributed over the surface of the sphere must appear somewhere.
Equations \eqref{s4} have not been studied yet. It is still unclear
whether they have at least one first integral and whether they are
Hamiltonian.

{\bf Acknowledgement.} We are grateful to M.\,A.\,Sokolovskiy for the reference to the work
\cite{bib2} and fruitful discussions. This work was supported in part by
CRDF (RU-M1-2583-MO-04), INTAS (04-80-7297), RFBR (04-205-264367) and NSh
(136.2003.1).

\newpage
\begin{figure}
\begin{center}
\includegraphics{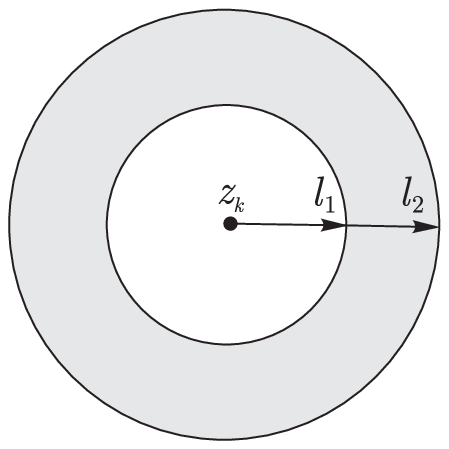}
\end{center}
\caption{}
\end{figure}

\begin{figure}[ht!]
\begin{center}
\includegraphics{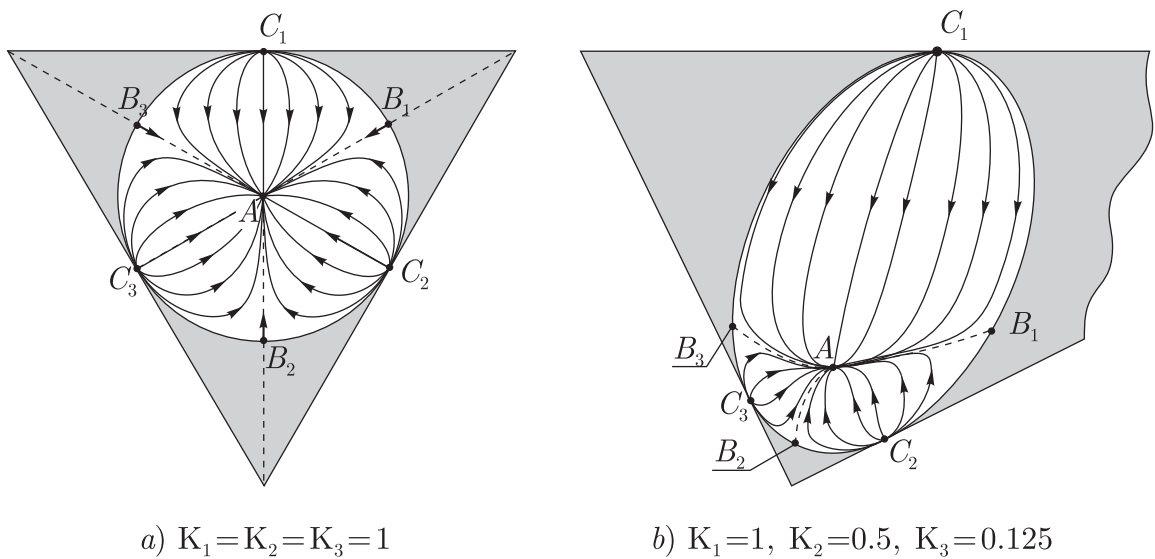}
\caption{Trajectories
of the system \eqref{g2} on the plane \eqref{g3} in the case 1$^\circ$.}
\end{center}
\end{figure}

\begin{figure}[ht!]
\begin{center}
\includegraphics{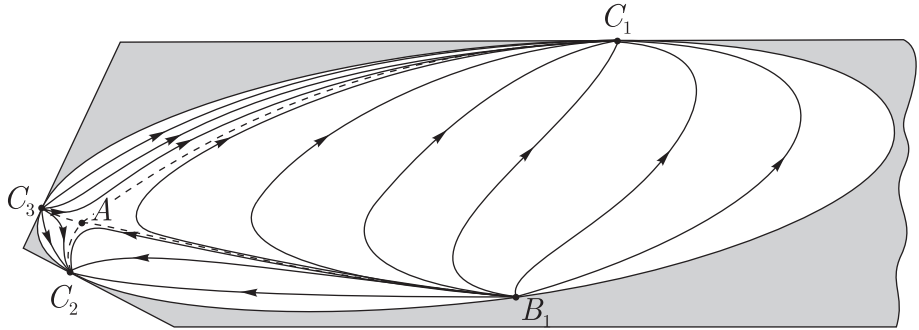}
\caption{Trajectories of the system \eqref{g2} on the plane \eqref{g3} in the case
$2^\circ$ ($\mathrm{K}_1 = 1$, $\mathrm{K}_2 = 0.5$, $\mathrm{K}_3 =
0.125$).}
\end{center}
\end{figure}

\begin{figure}[ht!]
\begin{center}
\includegraphics{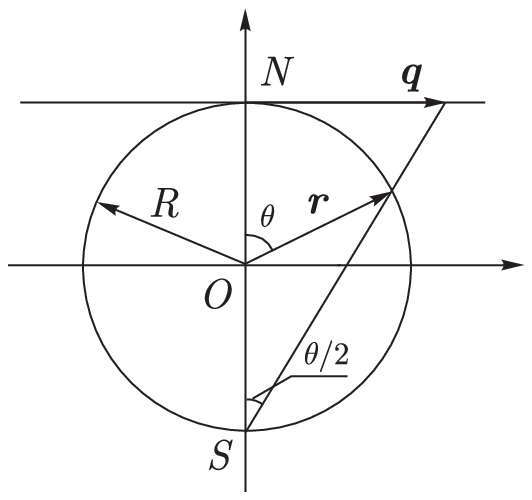}
\caption{}
\end{center}
\end{figure}

\newpage
Figure 1.

Figure 2. Trajectories
of the system \eqref{g2} on the plane \eqref{g3} in the case 1$^\circ$.

Figure 3. Trajectories of the system \eqref{g2} on the plane \eqref{g3} in the case
$2^\circ$ ($\mathrm{K}_1 = 1$, $\mathrm{K}_2 = 0.5$, $\mathrm{K}_3 =
0.125$).

Figure 4.


\newpage
\begin{thebibliography}{99}

\bibitem{bib5}
Newton P.\,K. {\em The $N$-Vortex problem: Analytical Techniques}, Springer,
2001.

\bibitem{bib1}
Fridman A.\,A., Polubarinova P.\,Ya. {\it On moving singularities of a
flat motion of an incompressible fluid.} Geofizicheskii sbornik, 1928,
p.~9--23.

\bibitem{bib2}
Bogomolov V.\,A. {\it Motion of an ideal constant-density fluid in the
presence of sinks.} Izv. Akad. Nauk SSSR. Mekhanika zhidkosti i gaza.
1976, No. ~4, p.~21--27.

\bibitem{bib8}
Jones S.\,W., Aref H. {\it Chaotic advection in pulsed source-sink
systems}. Phys. Fluids, 1988, v. 31(3), p. 469--485.

\bibitem{bib9}
Stremler M., Haselton F.\,R., Aref H. {\it Designing for chaos}\/: {\it
applications of chaotic advection at the microscale}. Phil. Trans. R. Soc.
Lond. A. 2004, v. 362, p. 1019--1036.

\bibitem{Kirh}
Kirchhoff~G. \emph{Mechanics. Lectures on mathematical physics.} Moscow:
Akad. Nauk SSSR, 1962. Transl. from German. Kirchhoff~G. \emph{Vorlesungen \"{u}ber
mathematische Physik}. Mechanik, Leipzig, 1874.

\bibitem{81}
Helmholtz~H. \emph{\"{U}ber Integrale hydrodinamischen Gleichungen weiche
den Wirbelbewegungen entsprechen}. J. Rein. Angew. Math., 1858, v.~55,
s.~25--55., see also Russian translation with Chaplygin's comments in the
book Helmholtz G. \emph{Fundamentals of vortex theory.} -- Moscow-Izhevsk: IKI,
2002.

\bibitem{bib3}
Borisov\,A.\,V., Mamaev\,I.\,S. {\it Poisson  structures and Lie algebras
in the Hamiltonian mechanics.} Moscow--Izhevsk: NITs RKhD, 1999.

\bibitem{bib7}
Borisov\,A.\,V., Mamaev\,I.\,S. {\it Mathematical methods of vortex
structure dynamics}. In ``Fundamental and applied problems of vortex
theory'' (A.V. Borisov, I.S. Mamaev, M.A. Sokolovskiy, Eds.) Collection of
papers --- Moscow-Izhevsk: IKI, 2003, p.~17-178.

\end{thebibliography}
\end{document}